\documentclass[aps,prc,reprint,
]{revtex4-2}

\usepackage[english]{babel}

\usepackage[letterpaper,top=2cm,bottom=2cm,left=3cm,right=3cm,marginparwidth=1.75cm]{geometry}

\usepackage{amsmath}
\usepackage{graphicx}
\usepackage{dcolumn}
\usepackage{bm}
\usepackage{braket}
\usepackage[colorlinks=true, allcolors=blue]{hyperref}
\usepackage{lineno}

\begin{document}

\title{Experimental Scheme for Polarizing the Boron Nuclei}

\author{William R. Milner}
\email{wmilner@mit.edu}
 \affiliation{Department of Physics, Massachusetts Institute of Technology, Cambridge, MA 02139
 }

\author{Richard G. Milner}
 \affiliation{Laboratory for Nuclear Science, Massachusetts Institute of Technology, Cambridge, MA 02139
 }

\preprint{APS/123-QED}

\date{\today}

\begin{abstract}

Unravelling the internal structure of hadrons and nuclei in terms of the quarks and gluons of Quantum Chromodynamics is a central focus of current nuclear physics research. Directly observing gluonic states in the nucleus would be groundbreaking and is an objective of the future Electron-Ion Collider (EIC). Over thirty years ago, Jaffe and Manohar identified a new double-helicity flip structure function, directly sensitive to exotic gluons. They pointed out that this could be measured in inclusive high-energy electron scattering from a transversely polarized nuclear target with spin $I \ge 1$.  Here, we identify the spin-3 nucleus boron-10 as a particularly interesting system to search for exotic gluons.  Leveraging technical advances in atomic physics over the past decade, we outline an experimental scheme to directly optically pump a beam of stable boron atoms to polarize the nuclear spin. Technical challenges to realize a spin-polarized beam of boron-10 in the EIC are discussed. The proposed scheme will also polarize the $^{11}$B nucleus, which could significantly enhance the pB fusion cross section. 

\end{abstract}
\maketitle
\section{Motivation}

A major thrust of twenty-first century nuclear physics is understanding how the quarks and gluons of Quantum Chromodynamics (QCD) give rise to the observed structure and properties of nuclei.  For example, the proton and neutron, the building blocks of nuclei, arise in QCD from light-mass quarks moving at relativistic speeds with enormous color interactions mediated via gluon exchange.
Major research efforts in experiment, theory and simulation directed at gaining increased insight are currently in progress worldwide.  A unique and innovative new accelerator, the Electron-Ion Collider (EIC), is being realized at Brookhaven National Laboratory, Upton, NY with the aim of opening up new windows on understanding QCD.

Despite the fundamental role the gluon ﬁlls in QCD, direct measures of gluonic states in the nucleus remain elusive. As the electrically neutral gluon does not couple directly to the photon, it is probed only indirectly in electron scattering from hadrons, but the dominance of the gluonic parton distribution function at low Bjorken $x$ highlights the importance of gluonic interactions in the nucleon. While to ﬁrst order nuclei are bound states of protons and neutrons, as the spin of the nucleus increases, higher-order behavior in the nucleus becomes available in the form of additional nuclear structure functions.

In 1989, Jaﬀe and Manohar identifed~\cite{Jaffe:1989xy} a new, leading-twist structure function which is sensitive to gluonic states in the nucleus but is free from contributions from the motion and binding of nucleons in the nucleus. This quantity, $\Delta (x, Q^2)$, is not sensitive to the contributions of bound nucleons or pions in the nucleus, as neither can contribute two units of helicity, and likewise neither can any state with spin less than one contribute. 
A measurement of $\Delta (x, Q^2)$ would require an unpolarized electron beam incident on a transversely 
polarized target of nuclear spin $I \geq 1$, and would constitute a pioneering search for previously unexplored {\it exotic gluons}.

The lowest moment of $\Delta(x,Q^2)$ has been estimated~\cite{Sather1990} in the MIT bag model.  Further, lattice QCD calculations of $\Delta(x,Q^2)$ have been carried out on the spin-1 $\phi(1019)$ vector meson~\cite{Detmold2016} and on the spin-1 deuteron~\cite{NPLQCD2017}.  In each case, a definitive signal for the new, double-helicity-flip structure function $\Delta(x,Q^2)$ was seen.

\begin{table}[h!]
    \centering
    \begin{tabular}{|l|c|c|c|c|}
    \hline
    {\bf Nucleus}&{\bf Nuclear}&{\bf Binding}& {\bf Spin} & {\bf Magnetic}\\
    & & {\bf Energy/} &{\bf per} &{\bf moment}\\
    &{\bf Spin} & {\bf nucleon} & {\bf nucleon}& $\boldsymbol \mu$\\
    \hline
    &{\bf I} & {\bf MeV} & & ${\boldsymbol \mu_{\bf N}}$\\
    \hline   
    \hline    
    $^2$H    & 1             & 1.1 & 0.50 & 0.86\\
    $^3$He   & $\frac{1}{2}$ & 2.6 & 0.17 & $-$2.13\\
    $^6$Li   & 1             & 5.0 & 0.17 & 0.82\\
    $^7$Li   & $\frac{3}{2}$ & 5.6 & 0.21 & 3.26\\
    $^9$Be   & $\frac{3}{2}$ & 6.7 & 0.17 & $-1.18$\\
    $^{10}$B & 3             & 6.5 & 0.30 & 1.80\\
    $^{11}$B & $\frac{3}{2}$ & 6.9 & 0.14 & $-$2.63\\
    $^{14}$N & 1             & 7.5 & 0.07 & 0.40 \\
    $^{17}$O & $\frac{5}{2}$ & 7.8 & 0.15 & $-1.89$\\
    $^{23}$Na & $\frac{3}{2}$& 8.1 & 0.07 & 2.22\\
    \hline
    \hline
    \end{tabular}
\caption{Binding energies and spin properties of some stable, light nuclei.}
    \label{tab:targ}
\end{table}

EIC~\cite{EICYRep} is an ideal facility to search for exotic gluons~\cite{Maxwell:2018gci,Maxwell2025}.  Both polarized electron and ion beams are planned in the design of the collider.  What is essential is a polarized ion beam with spin 1 or greater. Ideally, the nucleus should have significant binding, to enhance the amount of exotic gluons, but should not be too large, as spin effects will be diluted in the presence of a large number of unpolarized nucleons. That is, one wants to optimize the nuclear spin \textit{per nucleon}.  Table~\ref{tab:targ} summarizes the spin properties of a selection of light nuclei that have been considered. Magnetic moments are given in terms of nuclear magnetons ($\mu_N$). For comparison, the proton (neutron) magnetic moments are 2.79 ($-$1.93) $\mu_N$.   We note that $^{10}$B is unique in having a sizable binding energy and the largest spin/nucleon of any nucleus beyond the maximal value of the deuteron.

Boron has two stable isotopes: boron-10 ($I = 3$, 20\% abundance naturally) and boron-11 ($I = 3/2$, 80\% abundance naturally).  The Fermi momentum of the boron-10 nucleus is $\approx 220$ MeV/c and the total nuclear binding energy is $\approx 65$ MeV.  Boron-10 has five protons and five neutrons.  The nuclear spin of 3 for boron-10 results from the sum of angular momenta of the unpaired proton and neutron, each in the $1p_{3/2}$ shell model state. The addition of the extra neutron in boron-11 means the neutron contribution goes to zero and the nuclear spin of 3/2 results from the unpaired proton in the $1p_{3/2}$ state.

In this paper, we investigate the possibility of achieving a beam of polarized $^{10}$B ($I = 3$) nuclei. We present a scheme to directly optically pump $^{10}$B using lasers. Relying on the hyperfine interaction, circularly polarized laser beams transfer angular momentum to the atoms to polarize the nucleus. We outline the experimental techniques used to realize this scheme and discuss the technical challenges associated with achieving high polarization and using this beam in the EIC. 

\section{$^{10}\text{B}$ nuclear polarization scheme}\label{pumping_scheme}

The large nuclear spin of $I = 3$ for boron-10 makes it an attractive candidate for exotic gluon studies. However, the relatively complicated electronic structure of boron presents a challenge for optical pumping. Although beams of He-3 and alkali atoms have been optically pumped, they have a comparatively simple electronic structure. Alkali atoms are hydrogen-like with a single valence electron and spin-polarized beams of lithium~\cite{anderson1979proposal}, sodium~\cite{dreves1983production}, rubidium~\cite{jun1998pumping}, and cesium~\cite{masterson1993high} have been achieved.  Helium-3 possesses a metastable electronic state and established techniques for spin polarization have been developed using spin-exchange~\cite{abboud2004high}. Boron, on the other hand, is a ``group III" element, possessing a $p$-orbital electronic ground state $2s^2 2p \; ^{2}P_{1/2}$. Explorations into laser cooling and optical pumping of group-III elements have just begun in the past few years, with a beam of indium being successfully spin-polarized and laser-cooled~\cite{nicholson_MOT}. Investigations of laser cooling in Al~\cite{mcgowan1995light} and Ga~\cite{rehse2004laser} have also been demonstrated. Given that boron has a similar electronic level structure, the techniques used for indium can directly translate to boron. In addition, atomic beams of titanium~\cite{schrott2024atomic}, erbium~\cite{mcclelland2006laser}, dysprosium~\cite{youn2010dysprosium, lunden2020enhancing}, and chromium~\cite{chicireanu2006simultaneous} have been achieved, requiring ovens operating at high temperatures approaching $2000$ K and more complicated laser spectroscopy schemes than alkali atoms.

\begin{figure}[h!]
\center
\includegraphics[width=0.5\textwidth]{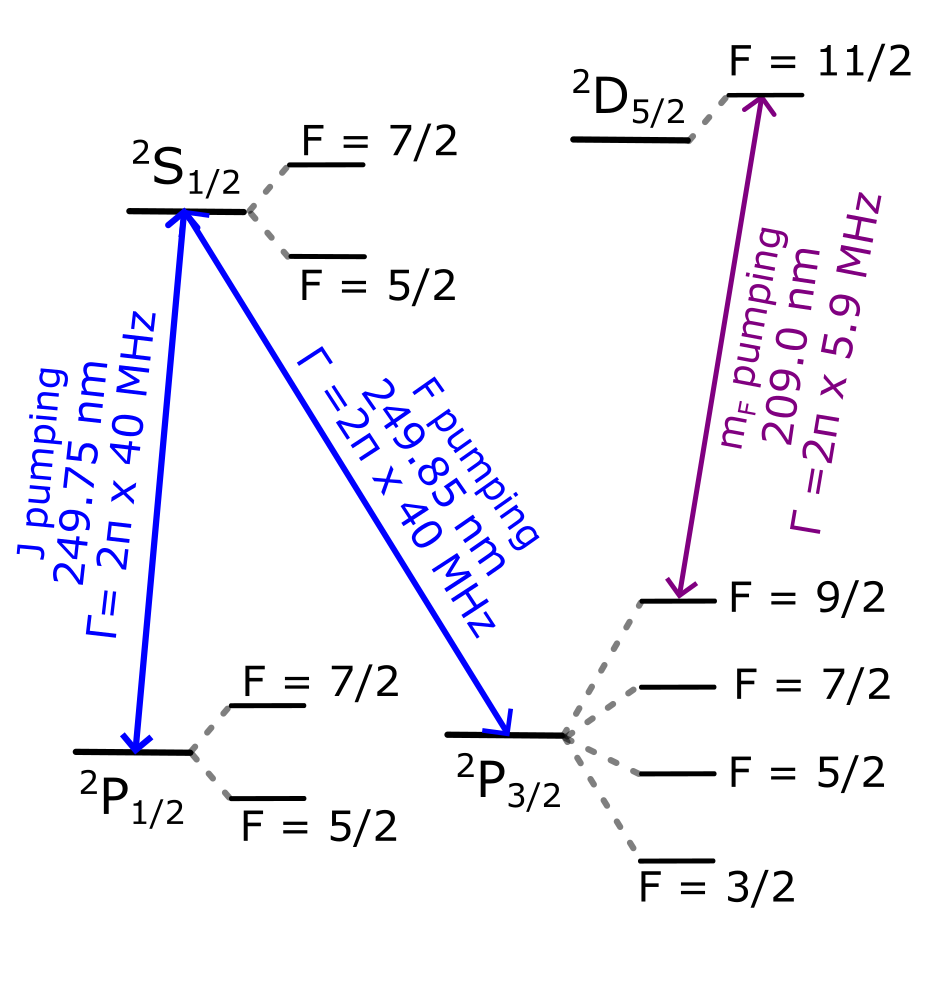}
\caption{$^{10}\text{B}$ energy levels. Transition rates $\Gamma$ are reported in Ref.~\cite{carlsson1994lifetimes}. Transition wavelengths are determined from Ref.~\cite{NIST_ASD}. Transitions are characterized in terms of $J$, $F$, and $m_{F}$ pumping steps. The $\ket{g}$ and $\ket{e}$ states are addressed by the $209$ nm transition.} 
\label{fig_1}
\end{figure}

\sloppy
All optical pumping schemes rely on the hyperfine interaction $H = A_{HF} I \cdot J$, which couples the electronic spin $J$ to the nuclear spin $I$. Before pumping, the atoms generally occupy a thermal distribution over all hyperfine sublevels $m_{F}$ $\{ -m_{F}$,$-m_{F} + 1$, $\dots$, $m_{F} \}$.  Using circular polarized light with angular momentum to drive electronic transitions, the atoms are pumped to a stretch state, $\ket{F, m_{F} = F}$ where the nucleus is fully polarized.  To effectively pump atoms, one wants to drive atoms to a closed transition where the excited state can only decay to a single, polarized ground state.  In boron, this condition can be achieved with the  $\ket{
2s^2 2p \;  ^{2}P_{3/2},F=9/2,m_{F}=9/2} \equiv \ket{g}$ to $\ket{2s 2p^2 \; ^{2}D_{5/2},F'=11/2,m_{F'}=11/2} \equiv \ket{e}$ transition. Due to selection rules, atoms in the $\ket{e}$ state can only decay to $\ket{g}$ and both states are fully nuclear polarized. We note that the state $2s 2p^2 \; ^{2}D_{J}$ lies energetically below $2s^2 3d \; ^{2}D_{J}$ for boron~\cite{NIST_ASD}, in contrast to In, Al, and Ga, and is the excited state used in  this scheme.

The energy levels of boron-10 are depicted in Fig.~\ref{fig_1}. Atoms start in an incoherent, thermal mixture in the $\ket{^{2}P_{1/2}}$ and $\ket{^{2}P_{3/2}}$ states. To transfer atoms to the target $\ket{g}$ state, the $J$, $F$, and $m_{F}$ quantum numbers must all be modified. First, atoms need to be transferred to the metastable $\ket{^{2}P_{3/2} }$ state. This is achieved using two `J pumping' lasers to drive both $\ket{^{2}P_{1/2},F= 5/2, 7/2}$ states to the $\ket{2s^2 3s\;^{2}S_{1/2},F'=7/2}$ state with a lifetime of $\tau \approx 4$ ns. To address all $m_{F}$ states, the lasers can be frequency modulated. Atoms rapidly decay and are distributed in the $F = 5/2, 7/2,$ and $9/2$ states in  the $J = 3/2$ manifold. Three additional lasers drive the `F pumping' transition $\ket{^{2}P_{3/2},F= 3/2, 5/2, 7/2}$ to $\ket{  ^{2}S_{1/2},F' =7/2}$  so that atoms accumulate in the $\ket{^{2} 
P_{3/2},F=9/2}$ manifold. All lasers should be turned on simultaneously to minimize de-pumping back to the $\ket{^{2}P_{1/2} }$ state. For the $J$ and $F$ pumping steps the polarization of the light is not critical, as the atoms are distributed among many $m_F$ states in $\ket{^{2}P_{3/2},F=9/2}$ following pumping. 

\begin{figure}[h!!]
\center
\includegraphics[width=0.5\textwidth]{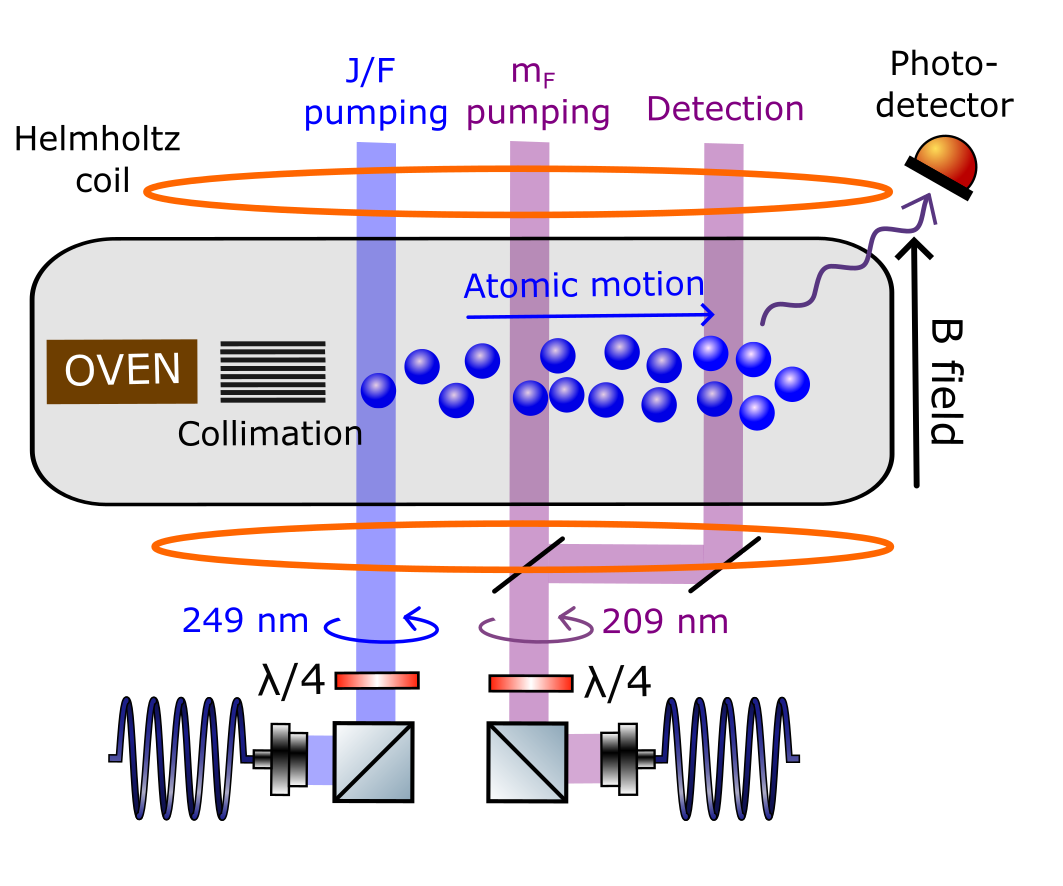}
\caption{Experimental schematic to prepare and measure spin polarization. Atoms are heated in an oven to $2000$ K, then collimated to reduce the transverse velocity spread. A pair of Helmholtz coils establishes a quantization axis transverse to the atomic beam motion. To avoid inhomogeneous fields, the coils should be large compared to the apparatus. $249$ nm light pumps atoms to the $\ket{^{2}P_{3/2},F=9/2}$ state. A secondary laser at $209$ nm optically pumps the atoms to the $\ket{^{2}P_{3/2},F=9/2, m_{F} = 9/2}$ state. A third spectroscopy region employs fluorescent detection to determine the state populations and thus nuclear polarization. }
\label{fig_2}
\end{figure}

With atoms in the target $\ket{^{2}P_{3/2},F=9/2}$ manifold, we now perform optical pumping (`$m_{F}$ pumping') to transfer atoms to a single, polarized $m_{F}$ state. As depicted in Fig.~\ref{fig_2}, circularly polarized light with handedness $\sigma^{\pm}$ transfers angular momentum so each absorbed photon drives a $\ket{m_{F}} \rightarrow \ket{m_{F'} = m_{F} \pm 1}$ transition. After absorbing many photons,  atoms are pumped to the $\ket{g}  \rightarrow \ket{e}$ cycling transition. Upon turning off the $m_{F}$ pumping light, any atoms in the $\ket{e}$ state quickly decay in $\tau =  27$ ns to the $\ket{g}$ state. 

\section{Technical realization}

{\it Thermal effects:} To experimentally realize the optical pumping scheme in Sec.~\ref{pumping_scheme}, there are a number of technical challenges that must be addressed. First and foremost, thermal effects of the atomic beam must be considered. The oven needs to operate at a temperature of $\approx 2000$ K to reach the same vapor pressure as Rb at room temperature.  To achieve these high temperatures, a high density graphite crucible could be used~\cite{ross1995high, maass_thesis}. Both boron-10 and boron-11 are available commercially in ultra-high purity as crystalline solids. For thin film applications, they are available as rod, pellets, pieces, granules and sputtering targets and as either an ingot or powder. 
At $2000$ K, the mean velocity $\langle v \rangle = \sqrt{8 k_{B} T / \pi m}\approx 2000$ m/s, corresponding to a rather large Doppler broadening of $\delta \omega = k \cdot v = 2 \pi \times 8.2$ GHz for the $249$ nm transition. This degree of broadening is substantially larger than the hyperfine splitting of $428.8$ MHz of the $\ket{^{2} P_{1/2}}$ states from Ref.~\cite{puchalski2015explicitly, maass_thesis} and makes the electronic states essentially unresolved at high temperature. We also note that given the rather small fine structure splitting of the $\ket{^{2}P_{1/2}}$ and  $\ket{^{2}P_{3/2}}$ 
 states~\cite{NIST_ASD} - $ h c \times 15.29\text{cm}^{-1} = k_{B} \times 22.0 \; \text{K}$ - the atoms at $2000$ K will occupy both $J = 1/2, 3/2$ states in the $\ket{^{2}P_{J}}$ manifold upon heating in the oven. 

The issue of thermal broadening can be strongly diminished by collimating the atomic beam to minimize the velocity spread in the transverse direction, so the transverse velocity is reduced to $v_{\perp} = \langle v \rangle \; \text{sin}(\theta)$.  To achieve a broadening equal to the $\ket{^{2}P_{1/2}}$ hyperfine splitting, a $3.0^{\circ}$ degree divergence half-angle would be necessary. Using a micronozzle oven scheme, a $1.2$ degree divergence half-angle was achieved in Ref.~\cite{senaratne2015effusive} using lithium with a similar mass. At the cost of reduced atomic flux, skimmers can be employed to further reduce the transverse beam spread. For example, using a molecular skimmer the transverse velocity of an atomic beam of lithium was reduced from 250 m/s to 0.1 m/s, achieving an optical pumping fidelity of $> 95 \%$~\cite{gillot2013optical}. Exploring buffer gas cooling to further reduce the transverse velocity could also be beneficial~\cite{hutzler2012buffer}.

{\it Nuclear polarization measurement:} For the $J$, $F$, and $m_{F}$ pumping steps which drive the atoms to the $\ket{g}$ state, resolving the individual $m_{F}$ states is not so important and the external magnetic field is just required for maintaining a quantization axis. For measuring the spin polarization via optical spectroscopy, the individual $m_{F}$ states must be resolvable. While the $m_{F}$ pumping must be done at low field to ensure a strong hyperfine coupling, the magnetic field can be adiabatically ramped to larger values for detection. In this section, we discuss spectroscopy in both the low and high magnetic field regimes. The  best resolution is likely achieved in the high field regime. 

\begin{figure}[h!!]
\center
\includegraphics[width=0.48\textwidth]{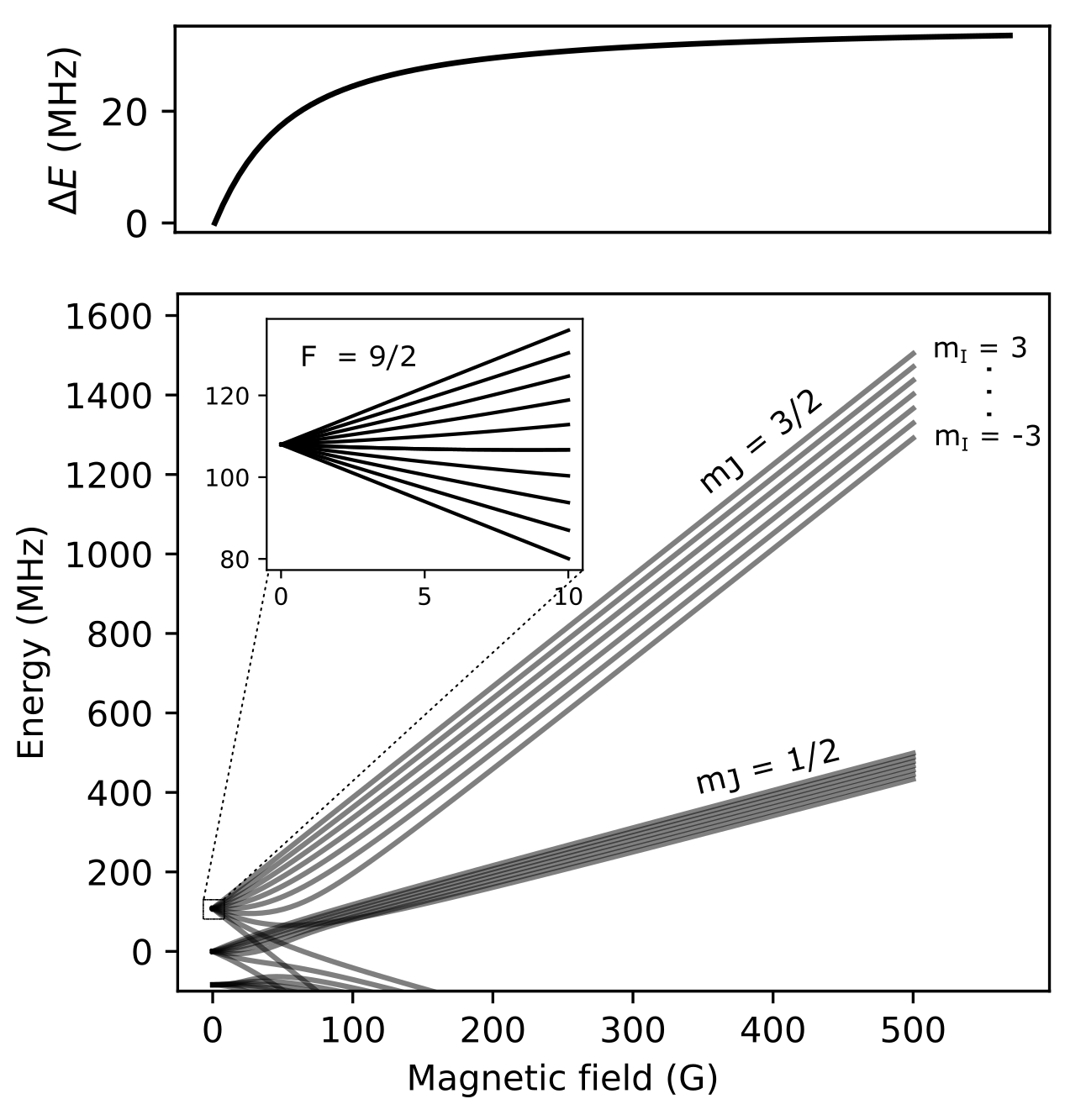}
\caption{$^{2}P_{3/2}$ energy levels. We diagonalize the Breit-Rabi Hamiltonian $H = A_{HF} I \cdot J + \mu_{B} g_{J} J_{z} B$ to determine the energy spectra. Lower panel: At low field, $F$ is a good quantum number and  $E = \mu_{B} g_{F} m_{F} B $. $m_{F}$ pumping to the $\ket{F = 9/2, m_{F} = 9/2}$ state will be employed in this regime. At high fields, the nuclear and electronic spin are decoupled into the $\ket{m_{J}, m_{I}}$ basis. In this regime, $E = A_{HF} m_{J} m_{I} + \mu_{B} g_{J}m_{J} B$ ignoring the much smaller electric quadrupole and nuclear g-factor terms. Nuclear polarization spectroscopy should likely be done in this regime to achieve the best resolution.
Upper panel: The splitting for the $\ket{F = 9/2, m_{F} = 7/2, 9/2}$ ($\ket{m_{J} = 3/2, m_{I} = 3, 2}$) states at low (high) magnetic field is plotted. While the splitting increases for larger magnetic fields, it saturates at $A_{HF} m_{J} = 37$ MHz for the $\ket{^{2}P_{3/2}}$ state. Although the energy levels for the $\ket{^{2} D_{5/2}, F' = 11/2}$ states will be qualitatively similar, the hyperfine constant $A_{HF'}$ required to directly calculate the energy structure is unknown to the best of our knowledge.} 
\label{fig_4}
\end{figure}

First, we discuss the low field regime where the energy splitting is less than the hyperfine structure. Here, $F$ is a good quantum number and the energy is well described by \\ 
$E = \mu_{B} B (g_{F'} m_{F'} - g_{F} m_{F})$, where
\begin{eqnarray}
g_{F} &=& g_{J} \frac{F(F +1) + J(J + 1) - I(I + 1)}{2F(F+1)} \ ,  \\
g_{J} &=& 1 + \frac{J(J +1) + S(S + 1) - L(L + 1)}{2J(J+1)} \ .
\end{eqnarray}
We calculate $g_{J} = 1.2$ and $1.33$ for $\ket{^{2}D_{5/2}}$ and  $\ket{^{2}P_{3/2}}$ states respectively, translating to magnetic sensitivities of $0.76$ and $0.62$ MHz/G/$m_{F}$ for the $\ket{^{2}D_{5/2}}$ and  $\ket{^{2}P_{3/2}}$ states. The differential magnetic sensitivity for $m_{F}$ states in the $\ket{g} \rightarrow \ket{e}$ transition is thus $1.4$ MHz/G, with sensitivies of $4.2$ and $2.8$ MHz/G for the $\ket{^{2}D_{5/2}}$  $m_{F'} = 11/2$ and $\ket{^{2}P_{3/2}}$ $m_{F} = 9/2$ states respectively. Although the hyperfine structure for the $\ket{^{2}P_{3/2}}$ state has not been measured, the A and B hyperfine coefficients were calculated in Ref~\cite{puchalski2015explicitly}, corresponding to  $112.1, -2.8, -86.9, -144.2$ MHz for the $F = 9/2, 7/2, 5/2, 3/2$ states respectively. This requires the B field to not exceed $40$ G to ensure the splitting is smaller than the hyperfine structure of the $F = 9/2$ and $F = 7/2$ states. The energy levels in the low field regime is plotted in the inset of the lower panel in Fig.~\ref{fig_4}. 

To spectroscopically determine the state populations to confirm the nuclear spin polarization, each $\sigma^{+}$ transition ($m_{F} \rightarrow m_{F'} = m_{F} + 1$) must be driven. The frequency splitting between individual spectroscopy transitions at $40$ G, using the calculated low field sensitivity, is $5.6$ MHz and nearly equivalent to the $6$ MHz linewidth of the $\ket{^{2} D_{5/2}}$ state.  Although this spectroscopy is not limited by the excited state linewidth, a $5$ MHz splitting places stringent requirements on the transverse velocity distribution. An $0.03^{\circ}$ divergence angle is required to achieve $5$ MHz thermal broadening on the $209$ nm transition. This motivates operating at high magnetic field, where the frequency splitting can be increased and the velocity requirements can be relaxed. Although Doppler-free spectroscopy techniques could be employed for detection, saturating the $209$ nm transition proves technically challenging and is explained later in the paper. Strictly speaking, at $40$ G the atoms are no longer in the low field limit and to more accurately calculate the transition splitting one should diagonalize the Breit-Rabi Hamiltonian as depicted in Fig.~\ref{fig_4}. However, as explained in the next section, this calculation requires the A hyperfine coefficient for the $\ket{^{2} D_{5/2}}$ state, which has not been measured or calculated for boron-10 to the best of our knowledge. 

At high field where the shifted energy levels far exceeds the hyperfine structure, the electronic and nuclear spin are decoupled and $F$ is no longer a good quantum number. In this regime, the states are characterized by the $\ket{m_{J}, m_{I}}$ basis and the energy levels converge at high field to:

\begin{equation}\label{hfeq}
    E = A_{HF} m_{J} m_{I} + \mu_{B} g_{J}m_{J} B. 
\end{equation}

Two lower contribution terms are the electric quadrupole, which is a percent level correction to the hyperfine structure, and the nuclear $g$-factor term, which is $1000$ times smaller than $\mu_{B} g_{J}m_{J} B$. As depicted in Fig.~\ref{fig_4}, at high field the energy levels are split into $m_{J}$ branches, where each branch contains a manifold of $2 I + 1 = 7$ states, each with energy $A_{HF} m_{J} m_{I}$.  

We first analyze the magnetic field dependence of the $\ket{^{2}P_{3/2}}$ state at high field. For this state, $A_{HF}$ was calculated in Ref~\cite{puchalski2015explicitly} to be $24.6$ MHz. Although the energy levels linearly increase according to the $\mu_{B} g_{J}m_{J} B$ term, the splitting between $m_{I}$ states saturates at $A_{HF} m_{J} \Delta m_{I}$. As seen in the top panel of Fig.~\ref{fig_4}, at the largest fields the splitting between the $\ket{m_{J} = 3/2, m_{I} = 3, 2}$ states approaches $A_{HF} m_{J}  = 37$ MHz.

Next we estimate the frequency splitting between  $\sigma^{+}$ transitions from $\ket{^{2}P_{3/2}, m_{J}, m_{I}}$ to $\ket{^{2}D_{5/2}, m_{J'}, m_{I}}$. For two neighboring transitions $\ket{^{2}P_{3/2}, 3/2, 3} \rightarrow \ket{^{2}D_{5/2}, 5/2, 3}$ and $\ket{^{2}P_{3/2}, 3/2, 2} \rightarrow \ket{^{2}D_{5/2}, 5/2, 2}$,  the frequency splitting between transitions in the high field limit is $A_{HF'} m_{J'} - A_{HF} m_{J}$ from Eq.~\ref{hfeq}. Intuitively, at high fields the differential sensitivity arises from the ground and excited state $A_{HF}$ terms with no magnetic field dependence. 

$A_{HF'}$ has not been measured or calculated for the $\ket{^{2}D_{5/2}}$ state. If $A_{HF'} = A_{HF}$, then the splitting between transitions would exactly be $A_{HF} = 24.6$ MHz, already exceeding the $5.6$ MHz separation at $40$ G, and corresponding to an $0.15^{\circ}$ beam divergence angle. Using the value for the $\ket{^{2}S_{1/2}}$ state from Ref.~\cite{maass_thesis}, $A_{HF'} = 79.3$ MHz, corresponds to a much larger $161$ MHz frequency splitting and approximately a $1^{\circ}$ atomic beam divergence angle.  Although the frequency splitting is generally larger in the high field limit, in the case that $A_{HF'} = A_{HF} m_{J}/m_{J'}$ the splitting is zero at high field and the optimum is at an intermediate magnetic field.

{\it Optical power constraints:} Due to limitations in optical power at ultraviolet wavelengths, the laser beams must be configured to scatter as many photons from the atomic beam as possible. Thus the laser intensity should be kept below saturation to ensure no optical power is wasted. The saturation intensity of the $\ket{g} \rightarrow \ket{e}$ transition is $\approx$ 100 mW/$\text{cm}^2$, which is relatively large given that the transition frequency is in the ultra-violet. $\sim1$ mW lasers are commercially available at $200$ nm. The vertical waist of the laser beams should be focused to match the typical atomic beam diameter of $1$ mm. 1 mW of power focused to a $1$ mm beam waist would achieve a scattering rate of $ \approx \Gamma / 4$. Given the longitudinal velocity of the atoms is 2000 m/s, an optical beamsize of 1 mm corresponds to an interrogation time of $\approx$ 500 ns. At $\Gamma/4$ intensity $\approx$ 5 photons would be scattered traversing the beam. This relatively low number of scattered photons will limit the nuclear polarization. Although increasing the laser beam waist along the axis of the beamline would linearly increase the interrogation time, the beam intensity is inversely proportional to this waist and the net number of photons scattered would be unchanged. In Ref.~\cite{schrott2024atomic}, the beam was folded multiple times to re-use the optical power. Using a Fabry-P\'{e}rot cavity to enhance the circulating beam power while expanding the beam size could also be a viable technique to address this concern.   

{\it Outlook:} Before attempting to achieve nuclear spin polarization, the first experimental steps should likely involve spectroscopically measuring the hyperfine structure for the $\ket{^{2}P_{3/2}}$ and $\ket{^{2}D_{5/2}}$ states. The hyperfine structure of the $\ket{^{2}P_{3/2}}$ state has not been experimentally verified, and the atomic structure (e.g. A and B hyperfine coefficients) for the $\ket{^{2}D_{5/2}}$ state has not even been calculated. With these hyperfine energy levels experimentally determined, the optical pumping could be modeled via a rate equation analysis to more quantitatively determine the number of scattered photons necessary to achieve high nuclear spin polarization. 

In summary, optical pumping of boron-10 appears feasible, although there are technical challenges to achieve high nuclear spin polarization that need to be experimentally tested. High oven temperature is required for atomic beam operation and the hyperfine coupling for the $\ket{^{2}D_{5/2}}$ states will likely dictate the atomic beam divergence angle required for detection. Finally, achieving a large beam flux ideally approaching $10^{15}\text{cm}^{-2} \text{s}^{-1}$ will be of importance to successfully use this atomic beam in the EIC. A peak flux of $\approx 5 \times 10^{10}$ $\text{cm}^{-2} \text{s}^{-1}$ was achieved using a similar optical pumping scheme with indium in Ref.~\cite{yu2022zeeman} incorporating a Zeeman slower. Despite these technical uncertainties, experimentally testing this scheme appears absolutely worthwhile to open the door to potentially studying nuclear spin in an entirely new regime. 






\section{Collider Implementation}
    
{\it Making an ion beam:}
To utilize the polarized nucleus in a collider, it is necessary to produce a polarized ion beam to inject into the accelerator.  Thus, the polarized neutral atoms must be ionized while maintaining nuclear polarization.  The technique being pursued for polarized $^3$He nuclei at EIC~\cite{Zelensky} is to inject the neutral, polarized atoms into an existing Electron Beam Ionization Source where they are ionized and subsequently injected into the accelerator.  Such a technique should also work for polarized boron nuclei but needs to be developed.
    
{\it Spin control and manipulation in EIC:}
The control and manipulation of the spins of polarized nuclei in the EIC accelerator chain is not trivial~\cite{Hock,Haixin2023,Meot2023}. To manipulate the atomic beam motion, while ensuring that any changes to the quantization axis established by magnetic fields are adiabatic,  so-called ``Siberian Snakes" are employed which are dipole magnets rolled into a helix. The Thomas-BMT equation~\cite{Thomas,BMT} describes the motion of a particle's spin-vector in a magnetic field in terms of the particle's anomolous magnetic moment $G$, where 
\begin{equation}
G \equiv \frac{g-2}{2} = \frac{\mu \cdot A}{2 \cdot Z \cdot I} - 1 \ ,
\end{equation}
and $\mu$, $A$, $Z$, and $I$ are the nuclear magnetic moment, atomic mass, atomic number and spin, respectively~\cite{Courant1997}.  The spin rotation angle $\phi$ for a snake is given by
\begin{equation}
\phi = 2\pi\sqrt{1+ \large [\frac{(G+1/\gamma)qB_0}{m \beta c |k|} \large]^2}
\end{equation}
where $q$ is the charge, $B_0$ is the magnetic field, $m$ is the mass, $\beta$c is the velocity, and $|k|=R\frac{2\pi}{\lambda}$, with $R$ being the helicity and $\lambda$ being the twist period.  Note that the AGS has two partial snakes, which rotate the spin less than 180$^\circ$.  The EIC will have six full snakes, each of which consists of four helical dipoles, which rotate the spin 180$^\circ$ for protons and helions. Other polarized light ions will need to use a partial snake configuration. For deuterons, and other similar light ions, the solenoid field from the experiment can provide a small spin rotation, and thus, a small spin-tune gap. For the case of deuterons, a modest tune jump can be used.  For the two stable boron isotopes, we have $G({}^{10}B) = -0.4$ and $G({}^{11}B) = 0.97$. To determine optimum performance in EIC, detailed studies must be carried out. 

{\it Polarimetry:}
Rapid, precise beam polarization measurements will be crucial for meeting the goals of the EIC physics program as the uncertainty in the polarization propagates directly into the uncertainty for relevant observables (asymmetries, etc.). In addition, polarimetry will play an essential role in developing the optimum running configuration for the accelerator. Polarimetry of high-energy polarized proton beams has been pioneered at RHIC via measurement of (i) the Coulomb-Nuclear Interference process using a thin ribbon carbon target passed across the beam~\cite{Kurita} and (ii) elastic scattering from a polarized atomic hydrogen jet target~\cite{Okada}.  Measurement of the polarization of a boron beam will require development of scattering processes with known analyzing powers.

{\it Theoretical Issues:}
Even with a highly polarized spin 3 $^{10} $B beam circulating in EIC, there are significant theoretical issues that must be addressed to extract observables directly sensitive to $\Delta(x,Q^2)$.  $\Delta$ is related to the $^{10}$B double-helicity-flip contribution to the  deep inelastic scattering cross section $\sigma(h,H,h',H')$, where $ (h,h' )/ (H,H' )$ are 
\newline ${\rm (virtual-photon)}/ ^{10}$B helicities (primed before and unprimed after scattering):
 \begin{eqnarray}
\sigma(h,H,h',H')&\propto& W(h,H,h',H') \\ 
&=& W_{\mu\nu}(H,H') \ell^{\mu\nu}(h,h') \,.
 \end{eqnarray}
Here, $W_{\mu\nu}(H,H')$, is the imaginary part of the forward amplitude for scattering electromagnetic currents $j_{\mu,\nu}$ from a spin-3 target in helicity space and $\ell$ is the photon helicity tensor $ \ell^{\mu\nu}(h,h') =\epsilon^{*\mu}(h' ) \epsilon^ \nu(h)$~\cite{Jaffe1989,Hoodbhoy1989,Jaffe:1989xy}  $\Delta(x,Q^2)$ is obtained by summing contributions with $h'=h\pm 2$ and $H'=H\mp 2$.  $\sigma(h,H,h',H')$ is invariant under boosts along the helicity quantization axis, so the target density matrix can be measured in any frame collinear with the virtual photon, including the virtual photon-target center-of-mass. In the Bjorken limit, the result can be expressed in terms of the Lorentz invariant variables $x$ and $Q^2$.  This program has been carried out for spin-1~\cite{Jaffe:1989xy}, but will be significantly more complicated for a spin-3 target~\cite{Jaffe1995}.    For example, the target helicity space is 7-dimensional.  A further complication is that given the incoherent optical pumping scheme, the target will be in a mixture of helicity states (i.e. a mixed quantum state) at the point of collision. Using experimental polarimetry measurements, the distribution of helicity states can be directly determined.  	


{\it Polarizing the boron-11 nucleus:}
The electronic levels in Fig.~\ref{fig_1} are identical for the $^{11}$B atom, the other stable boron isotope.  A spin-3/2 boron-11 beam in EIC can be used to complement the spin-3 boron-10 beam in searches for exotic nuclear effects.
Further, this nucleus is also interesting for the aneutronic fusion process~\cite{Magee}
\begin{equation}
    {}^1{\rm H} + {}^{11}{\rm B} \rightarrow 3 \times  {}^4{\rm He} + 8.7\ {\rm MeV}\ , 
\end{equation}
    where the three $\alpha$ particles have an energy distribution up to 5 MeV.  The reactants, hydrogen and boron, are abundant in nature, non-toxic and nonradioactive, and the reaction itself produces no neutrons, only helium in the form of three alpha particles. There are challenges, most notably that the temperature required for a thermonuclear p$^{11}$B (pB) fusion reactor is 30 times higher than that for deuterium-tritium (DT), the candidate fusion fuel with the lowest operating temperature.  However, the spin-dependent pB fusion process has been calculated~\cite{Ahmed} and a substantial increase in the cross section is reported under optimal conditions, e.g. an increase of 60\% for the $^{11}$B polarized in the $+\frac{3}{2}$ magnetic substate.  Thus, the polarization scheme presented here may be used to enhance the pB fusion cross section.

\medskip
\section{Summary}
Recent advances in laser cooling and optical pumping of group-III atoms allow us to propose a scheme to produce nuclear polarization of the two stable boron isotopes.  Atomic beams generated from a hot oven can be optically pumped by available lasers to possibly produce high nuclear polarization.  Such a scheme can form the basis of a polarized boron-10 ion source for use at the future Electron-Ion Collider to search for exotic gluon states in nuclei. Issues for realization include: generating a sufficiently high flux, making a polarized ion beam, spin control and manipulation in EIC and determining the polarization of the stored boron-10 beam. Our proposed scheme can also be used to generate a polarized internal gas target of boron nuclei for use in a storage ring with recirculating beams.  Clearly, the first step is to construct a prototype source which can demonstrate feasibility of the atomic polarization process. Demonstrating a nuclear polarization $> 50 \%$ would already be potentially viable to use for gluon studies. The proposed scheme can also be used to produce polarized boron-11 nuclei, which can be used to enhance the pB fusion cross section.

\medskip
\section*{Acknowledgements}
We thank the organizers of the meeting on {\href{https://indico.cfnssbu.physics.sunysb.edu/event/343/}{Polarized Ion Sources and Beams at EIC}} at the {\href{https://www.stonybrook.edu/cfns/}{Center for Frontiers in Nuclear Science}} at Stony Brook University in March 2025 for the opportunity to initially present this work. We acknowledge significant contributions from R. Jaffe on identifying the theoretical issues, from K. Hock on understanding the spin control and manipulation in EIC, and from W. Ketterle on the optical detection scheme.
We thank Kyungtae Kim, Yu-Kun Lu, and James Maxwell for critical reading of our manuscript and Hanzhen Lin for sharing Breit-Rabi calculations. W.R.M acknowledges funding from the MIT Physics Department in Wolfgang Ketterle's group. R.G.M. is supported at MIT-LNS by the U.S. Department of Energy, Office of Nuclear Physics, under grant number DE-FG02-94ER40818.

\bibliographystyle{unsrt}
\bibliography{sample}

\end{document}